\documentclass{elsarticle}
\usepackage[utf8]{inputenc}
\usepackage{amsmath}
\usepackage{amssymb}
\usepackage{graphicx}
\usepackage{url}

\usepackage{color}
\def\beq{\begin{equation}}
\def\beqn{\begin{eqnarray}}
\def\eeq{\end{equation}}
\def\eeqn{\end{eqnarray}}

\newcommand{\sineff}{\sin^2 \theta^{{\rm lep}}_{{\rm eff}}}
\newcommand\sss{\scriptscriptstyle}
\newcommand{\tril}{\lambda_{3}}
\newcommand{\trilsm}{\tril^{\rm SM}}
\newcommand{\mh}{m_{ \sss H}}
\newcommand{\mw}{m_{ \sss W}}
\newcommand{\mz}{m_{ \sss Z}}
\newcommand{\ktre}{\kappa_{\lambda}}

\title{Higgs boson self-coupling constraints from single Higgs, double Higgs and Electroweak measurements}

\author[1]{Giuseppe Degrassi}
\author[1]{Biagio Di Micco}
\author[2]{Pier Paolo Giardino}
\author[3]{Eleonora Rossi$^1$}
\address[1] {Universit\`a degli Studi di Roma Tre, INFN  sezione di Roma Tre, I-00146 Roma, Italy }
\address[2]{Instituto Galego de F\'isica de Altas Enerx\'ias, Universidade de Santiago de Compostela, 15782 Santiago de Compostela, Galicia, Spain}
\address[3]{LAPP, Universit\'e Grenoble Alpes, Universit\'e Savoie Mont Blanc, CNRS/IN2P3, Annecy; France.}
\date{January 2021}

\begin{document}
\begin{abstract}
  We set constraints on the trilinear Higgs boson self-coupling, $\tril$, by
  combining the information coming from the $W$ mass and
  leptonic effective Weinberg angle, electroweak precision observables, with
  the single Higgs boson
  analyses targeting the $\gamma \gamma,\, ZZ^*,\, WW^*, \,\tau^+
  \tau^-$ and $\bar{b} b$ decay channels and  the double Higgs
  boson analyses in the $b\bar{b}b\bar{b}, \, b\bar{b}b \tau^+ \tau^-$ and
  $b\bar{b}b \gamma \gamma$ decay channels, performed by the ATLAS collaboration.
  With the assumption that the
  new physics affects only the Higgs potential, values outside the interval
  \mbox{$ -1.8\, \trilsm < \tril < 9.2 \, \trilsm$} are excluded at
  $95\%$ confidence level. With respect to similar analyses 
  that do not
  include the information coming from the electroweak precision observables our
  analysis shows a stronger constraint on both positive and negative values
  of $\tril$. 
\end{abstract}

\maketitle 
\footnotetext[1]{Corresponding author: eleonora.rossi@cern.ch}
\section{Introduction}

With the discovery of the Higgs boson \cite{Higgs,Brout,ATLAS,CMS},
the study of the Higgs boson potential and of the Higgs
self-interactions \cite{Spira} has become of great interest in the
scientific community \cite{HHWhitePaper}. The shape of the Higgs boson
energy potential and the value of the Higgs self-couplings have deep implications on cosmology
\cite{Masina,Strumia,Shaposchnikov} and on quantum field theory, in
particular in connection with gravity \cite{Shaposhnikov_asympt}.

In the Standard Model (SM) the coupling of the quartic term in the Higgs
potential,
$\lambda$, dictates the trilinear and quadrilinear Higgs self-interactions
and its value is related  to the Higgs
field vacuum expectation value, $v$, and the Higgs boson mass, $\mh$, by
$\lambda \equiv\trilsm= \mh^2/2v^2$. Thus, it  can be expressed as a
function of physically measurable quantities in terms of $G_F$ and
$\mh$ as $\trilsm = G_F \mh^2 \,\sqrt{2}$,  where $G_F$ is the Fermi coupling
constant, linked to $v$ via  $v =({\sqrt{2}\,G_F})^{-1/2}$,
whose value is obtained from the muon lifetime measurement: $G_F =
1.1663788 \times 10^{-5} \, \textrm{GeV}^{-2}$
\cite{MulanCollaboration}, while  $\mh$ is the Higgs boson mass measured
from the Higgs boson decay products \cite{PDG,ATLAS_13, CMS_13,
  ATLAS_CMS_H_MASS}, $\mh = 125.14$ GeV.

The Higgs self-interactions affect any observable either at the tree-level or
via quantum corrections. In particular, the trilinear Higgs coupling,
$\tril$, 
affects the double-Higgs boson  production, $pp \to HH$ at the tree-level
\cite{Spira, Maltoni} while both single-Higgs boson production
and decay processes are affected   at the one-loop level \cite{Degrassi_singleH}. 
Going on in the perturbative expansion, i.e. at the two-loop level,
$\tril$ affects observables with no Higgs bosons as external states, like
the electroweak precision observables (EWPO), in particular the $W$ boson mass,
$\mw$, and the leptonic
effective Weinberg angle $\sineff$ \cite{Degrassi_mW}. The latter differs
from the Weinberg angle $\theta_W$ defined in terms of the physical $W$ and
$Z$ boson masses through the relation $\cos \theta_W = \mw/\mz$, by a
renormalization factor $\kappa^{lep}$ such that
$\sineff = \kappa^{lep} \sin^2\theta_W$, where $\kappa^{lep}$ includes all higher
order corrections affecting the coupling of the Z boson to leptons.

While the couplings of the Higgs boson to vector bosons and fermions
have been measured at the $5\%$ level \cite{SingleHiggsComb,ATLAS_comb_2020, CMS_comb_2020},
among the Higgs self-interactions  only the trilinear one can be constrained
experimentally, although very weakly. Then, it is worth to use all the available
experimental information in order to strengthen the constraint on $\tril$,
albeit under some assumptions. At present the  constraint on
$\tril$ obtained by the ATLAS collaboration  combines the information from double 
 Higgs analyses with an integrated luminosity up to 
$36.1\: {\rm fb}^{-1}$  with  single Higgs analyses up to $79.8 \:{\rm fb}^{-1}$ 
reporting that values of $\tril$
outside the interval \mbox{$ -2.3\, \trilsm < \tril < 10.3 \, \trilsm$}
are excluded at $95\%$ confidence level (CL) \cite{ATLAS_H_HH_comb}. Instead the  CMS collaboration, using the process $pp \to HH \to b\bar{b}\gamma \gamma$ with an integrated luminosity of $137 \:{\rm fb}^{-1}$,  is able to exclude $\tril$ values  outside \mbox{$ -3.3\, \trilsm < \tril < 8.5 \, \trilsm$} 
at 95 \% CL  \cite{CMS_bbyy}.

In this letter we make a further step on the path of strengthening the
constraint on $\tril$ by combining the public information available
on the double and single-Higgs processes from the  ATLAS Collaboration
with the information coming from the EWPO. We perform a fit
to double and single-Higgs production cross sections and Higgs decay
channels together with the value of $W$ mass and $\sineff$ building a
likelihood function of one parameter of interest, $\ktre$, that measures the
deformation of the Higgs trilinear coupling with respect to its SM value,
or $\tril = \ktre \trilsm$.

The paper is organized as follows. In section \ref{sec:theory} we discuss
the theoretical framework in which our analysis is inserted. In section
\ref{sec:data} the experimental inputs that enter in our analysis are presented
and discussed. Section \ref{sec:fit} contains the fit procedure we employed,
while the next section contains the results of the various fits we perform.
Finally we present our conclusions.

\section{Theoretical framework}
\label{sec:theory}

In this section we briefly review some of the results presented in
refs.\cite{Degrassi_singleH,Degrassi_mW,Degrassi:2019yix} that were used as a
basis for our analysis.  We are interested in studying a Beyond the
Standard Model (BSM) scenario where the dominant effect of an unknown
new physics (NP) is concentrated on the modification of the Higgs potential
\begin{equation}
V(H)_{\rm BSM}=\frac{1}{2}\mh^2 H^2+\ktre \trilsm v H^3+\kappa_{\lambda 4}
\frac{\trilsm}{4} H^4+\cdots,
\label{eq:VH}
\end{equation}
where the dots represent higher orders interactions, while at the same time
the effects of NP on the other SM couplings are assumed to be negligible.
This scenario can be described by a Lagrangian that differs from the SM one
only in 
the scalar potential part that is modified via an (in)finite tower of
$( \Phi^\dagger \Phi )^n$ terms or 
\beq
V^{NP} = \sum_{n=1}^N \mathcal{C}_{2n} ( \Phi^\dagger \Phi )^n\,,  \qquad\qquad 
\Phi = \binom{0}{\frac{1}{\sqrt{2}}( v+ H)}\,,
\label{eq:VNP}
\eeq
with $\Phi$ the Higgs doublet, as shown in the Unitary gauge.

The $\kappa$ factors in eq.(\ref{eq:VH}) can be  easily related to the
coefficient $\mathcal{C}_{2n}$ in eq.(\ref{eq:VNP}). In particular
for the trilinear Higgs self-interaction one finds \cite{Degrassi_mW}
\beq
\ktre = 1 + 2 \frac{v^2}{\mh^2}\,
\frac13 \sum_{n=3}^N \mathcal{C}_{2n} \, n(n-1)(n-2)
\left(\frac{v^2}{2}\right)^{n-2} \,.
\label{dc3}
\eeq
 We remark that
the potential in eq.(\ref{eq:VNP}) is assumed to be general, i.e. the  coefficients $\mathcal{C}_{2n}$ are not supposed
to obey a hierarchy  scaling as
$\mathcal{C}_{2n+2} \sim \mathcal{C}_{2n} v^2/\Lambda_{NP}^2$, with $\Lambda_{NP}$
the scale of NP, like in a well-behaved Effective Field Theory (EFT).

The effects induced on the observables by a modified trilinear Higgs
self-interaction occur at different orders in the perturbative expansion
(tree or loop level) depending on the observable under consideration. When
these effects  appear for the first time at the loop
level, i.e. in single-Higgs processes and EWPO, 
the modification of the  observable induced by the lowest-order
contribution  can be parametrized as
\beq
\mathcal{O}_{\rm BSM}=\mathcal{O}_{\rm SM}\, \left( 1+(\ktre-1)C_1+
(\kappa_\lambda^2-1)C_2 \right),
\eeq
where $\mathcal{O}$ is a generic observable defined in the BSM scenario or
in the SM respectively, and $C_1$ and $C_2$  are finite numerical coefficients,
i.e. their values do not depend on $\Lambda_{NP}$.
\begin{table}[t]
\begin{center}
\begin{tabular}{c|r|r|}
      & $C_1$~~~~~ & $C_2$~~~~~ \\ 
\hline
$\mw$ & $ 5.62\times 10^{-6}$  & $-1.54 \times 10^{-6}$ \\
$\sineff$ & $ -1.56 \times 10^{-5}$ & $4.55  \times 10^{-6}$
\end{tabular}
\end{center}
\caption{ Values of the coefficients $C_1$ and $C_2$ for the EWPO.}
\label{tab:1}
\end{table}

The values of the $C_1,\, C_2$ coefficients for single-Higgs observables are
reported in ref.\cite{Degrassi_singleH} while those for the EWPO can be found in ref.\cite{Degrassi_mW}. Here 
we use the latest SM theoretical predictions for $\mw$ \cite{EWK_test} and $\sineff$ \cite{Dubovyk:2019szj} to refine the calculation of the latter coefficients. We employ as SM predictions  \mbox{$\mw = 80.359 \pm 0.06 $} GeV  and 
\mbox{$\sineff = 0.23151 \pm 0.00006$} where  the errors reported are obtained combining in quadrature the
parametric uncertainties with our estimate of the missing higher order terms~\cite{Degrassi_mW}.
In Table \ref{tab:1} the updated values of the $C_1,\, C_2$ coefficients are presented.

Before concluding this section we want to comment on the total uncertainties
that affect our analysis. For any measurement,  besides the  experimental
uncertainty, we  take into account a theory uncertainty that can be
divided in a part related to the SM prediction and another $\ktre$-dependent
associated to missing higher order terms. The former is usually already included in
the experimental analyses while the latter is actually very difficult to
estimate. In ref.\cite{Degrassi_singleH} the $\ktre$-dependent uncertainty
was estimated in terms of the process-dependent coefficient $C_1$, however
the result of that analysis showed a very mild dependence on this uncertainty.
Concerning the $\ktre$-dependent uncertainty of the EWPO we used the same kind
of estimate of ref.\cite{Degrassi_singleH} finding also in our case a very mild dependence on this uncertainty. 

\section{Data inputs}
\label{sec:data}
In this Section we discuss the experimental inputs we use in
the fit. The observables we consider are:
the \mbox{$pp \to HH \to b\bar{b} \gamma \gamma$}, the
\mbox{$pp \to HH \to b\bar{b} b\bar{b}$} and the
\mbox{$pp \to HH \to b \bar{b} \tau^+ \tau^-$}
production cross sections as measured
by the ATLAS collaboration \cite{ATLAS_HH_comb, ATLAS_4b, ATLAS_bbyy,
  ATLAS_bbtautau}; the measurements of the single-Higgs boson
production cross sections including the gluon fusion (ggF),
the vector boson fusion (VBF), the associate production (VH) and the $t\bar{t}H$
production modes;  the branching fractions of the
\mbox{$H \to \gamma \gamma$},
\mbox{$H \to ZZ$}, \mbox{$H \to W^+W^-$}, \mbox{$H \to b \bar{b}$} and
\mbox{$H \to \tau^+ \tau^-$} decay channels \cite{SingleHiggsComb};
the value of the
$W$ boson mass from the world average \cite{PDG, ATLAS_mW, ALEPH_mW,
  L3_mW, OPAL_mW, DELPHI_mW, CDF_mW, D0_mW};  $\sineff$
as estimated in ref.\cite{EWK_test} from the
average of the LEP \cite{ALEPH_sin, DELPHI_sin, L3_sin,
  OPAL_sin}, SLD \cite{SLD,LEP_SLD_comb}, Tevatron
\cite{DZERO_sin,CDF_sin,Tevatron_sin} and LHC \cite{CMS_sin, ATLAS_sin,
   LHCb_sin} data.

The $\sineff$ measurements are slightly
inconsistent due to a discrepancy at the level of 3$\sigma$ between
the LEP and the SLD most accurate measurements, namely the
measurement obtained from the forward-backward asymmetry
in the \mbox{$e^+e^- \to Z \to b \bar{b}$} at LEP and  the one obtained from
the left-right asymmetry A$_{\rm LR}$ in \mbox{$e^+ e^- \to Z \to l^+ l^-$} at
SLD. The $\chi^2$ of the fit is 11.5 with 5 degrees of freedom. In
order to not underestimate the error on the average, from combining
discrepant measurements, and to be conservative, we assume that the
discrepancy is due to an underestimated systematic error that affects
all measurements. Therefore all measurement uncertainties are
multiplied by a scaling factor $\kappa = \sqrt{11.5/5}$ such that the
$\chi^2$ of the fit of the combined measurements equals its
expectation value, this in turn consists in multiplying by 1.52 the error of
the average computed in ref.\cite{EWK_test}:
$\sineff = 0.23151 \pm 0.00014$.
The single-Higgs boson measurements were taken from the ATLAS
collaboration results, nevertheless few measurements were excluded
from the fit to extract $\ktre$. In particular the
$t\bar{t}H$ production mode with the $H \to \gamma \gamma$ decay mode
was excluded to avoid double counting between this channel and the $pp
\to HH \to b \bar{b} \gamma \gamma$ channel, as discussed in
ref.\cite{ATLAS_H_HH_comb}.  Table \ref{tab:input_summary} summarises
all the input measurements used in this work.

\begin{table}[htb]
\begin{center}
\begin{tabular}{lcc}
\hline
\multicolumn{3}{c}{Double Higgs-boson production (ATLAS data)} \\
\hline
\multicolumn{2}{l}{Channel} & $\mathcal{L} \,  $[fb$^{-1}$] \\
\hline
\multicolumn{2}{l}{$pp \to HH \to b\bar{b} \gamma \gamma$} & 36.1 \\
\multicolumn{2}{l}{$pp \to HH \to b\bar{b} b\bar{b}$} & 27.5 \\
\multicolumn{2}{l}{$pp \to HH \to b\bar{b} \tau^+ \tau^-$} & 36.1 \\
\hline
\multicolumn{3}{c}{Single Higgs-boson production (ATLAS data)} \\
\hline
Decay Channel & Production Mode & $\mathcal{L}$ [fb$^{-1}$] \\
\hline
$H \to \gamma \gamma$ & ggF, VBF, $WH$, $ZH$ & 139 \\
$H \to ZZ^*$ & ggF, VBF, $WH$, $ZH$, $t \bar{t}H$ & 36.1 - 139 \\
$H \to W^+ W^-$ & ggF, VBF, $t\bar{t}H$ & 36.1 \\
$H \to \tau^+ \tau^-$ & ggF, VBF, $t\bar{t}H$ & 36.1 \\
$H \to b \bar{b}$ & VBF, $WH$, $ZH$, $t\bar{t}H$ & 24.5 - 139 
\\
\hline
\multicolumn{3}{c}{Precision electroweak  observables} \\
\hline
Observable & Value & Reference \\
\hline
$m_W$ & $80.379 \pm 0.012$ GeV & PDG World Average \\
$\sin^2\theta_{\rm eff}^{lep}$ & $ 0.23151 \pm 0.00021$ & LEP/SLD/Tevatron/LHC \\
\hline
\end{tabular}
\end{center}
\caption{Input measurements used in the present work. For the ATLAS
  measurements the analysed dataset has been specified being the
  experiment still on-going and analyses updates on larger datasets
  are expected in the future.}
\label{tab:input_summary}
\end{table}

For the single-Higgs boson production and decay modes, it is
conventional to fit data using the production cross section and decay
branching fraction signal strengths ($\mu_i$,$\mu_f$), defined as the
ratio between the observed values and their SM expectations:
\beqn
\mu_{i} & = &\sigma^{\rm observed}_i/\sigma^{\rm SM}_{i}\, ,~~~~~~~~~~~
i = \mathrm{ggF}, \, \mathrm{VBF}, \, WH, \, ZH, \, t\bar{t}H \nonumber\\
\mu_{f} &= &\mathrm{Br}_{H \to f}^{\rm observed}/\mathrm{Br}_{H \to f}^{\rm SM}\, ,~~~~~
f = \gamma \gamma, \, ZZ^*, W^+W^-, \, b\bar{b}, \, \tau^+ \tau^- ~.
\nonumber
\eeqn

In this work the signal strengths are taken from
ref.\cite{SingleHiggsComb} where the product $\mu_i \times \mu_f$ is
tabulated for each production and decay mode; the values used in this
fit are summarised in Table \ref{tab:mus}.  The fit performed in
ref.\cite{SingleHiggsComb} combines the $ZH$ and the $WH$ channel
assuming that the ratio of their cross section equals its SM expectation.
Such assumption cannot be made in our case because
 $\ktre \neq 1$  affects differently the
$WH$ and $ZH$ cross sections. Nevertheless the sensitivity of the $VH$
result is dominated by the $H \to b \bar{b}$ channel where $ZH$
provides the most accurate measurement of the $VH$ signal strength
value \cite{Hbb}.
In this work the $VH$ signal strength is therefore assigned to $ZH$,
the impact of this assumption has been tested assigning the $VH$
signal strength of the $\gamma \gamma$ channel (the second most
relevant channel after $H \to b \bar{b}$) to $WH$ and decoupling the
$WH$ and $ZH$ signal strengths in the $H \to b \bar{b}$ channel using
inputs from ref.\cite{Hbb}. The impact on the result has been found
to be negligible.  In addition, the signal strength relative to
$t\bar{t}H + tH$ of the original paper has been assigned to
$t\bar{t}H$ being the $tH$ contribution negligible, and the $VV$
channel indicated in ref.\cite{SingleHiggsComb} has been assigned to
$W^+ W^-$ that dominates the sensitivity.  Finally the uncertainties
on the signal strengths have been symmetrised by averaging the squares
of the positive and negative uncertainties.

\begin{table}
\begin{center}
\begin{tabular}{c|ccccc}
  $\mu_i \times \mu_f$ & ggF & VBF & $ZH$ & $t\bar{t}H$ \\
  \hline
  $\gamma \gamma$ & 1.03 $\pm$ 0.11 &  1.31 $\pm$ 0.25 & 1.32 $\pm$ 0.32 & -- \\
  $ZZ^*$ & 0.94 $\pm$ 0.11 & 1.25 $\pm$ 0.46 & 1.53 $\pm$ 1.03 & -- \\
  $W^+W^-$ & 1.08 $\pm$ 0.19 & 0.60 $\pm$ 0.35 & -- & 1.72 $\pm$ 0.55 \\
  $b\bar{b}$ & -- & 3.03 $\pm$ 1.65 & 1.02 $\pm$ 0.18 & 0.79 $\pm$ 0.60 &  \\
  $\tau^+ \tau^-$ &  1.02 $\pm$ 0.58 & 1.15 $\pm$ 0.55 & -- & 1.20 $\pm$ 1.00\\
  \end{tabular}
    \caption{Values of the product $\mu_i \times \mu_f$ used in the
      fit procedure. The values corresponds to the product $\sigma
      \times \mathrm{Br}$ normalised to its Standard Model
      expectation. Rows span over the decay mode $f$ while colums over
      the production mode $i$.}
    \label{tab:mus}
    \end{center}
\end{table}
The measurements shown in Table \ref{tab:mus} are correlated due to
the cross contamination of signal events among different production
and decay channels and the correlation matrix is provided in Figure 6
of ref.\cite{SingleHiggsComb}; in the present work only correlations
larger than 0.05 are taken into account for simplicity; this choice
doesn't have impact on the final results. The correlation coefficients
are shown in Table \ref{tab:correlations}.

\begin{table}[t]
\begin{tabular}{c|c|cccc|c|cc}
& ggF & \multicolumn{4}{c|}{VBF} & $VH$ & \multicolumn{2}{|c}{$t\bar{t}H$} \\  
\hline
 $\rho$  & $\tau^+ \tau^-$ & $\gamma \gamma$ & $ZZ^*$ & $W^+W^-$ & $\tau^+ \tau^-$ & $ZZ^*$ & $W^+ W^-$ & $\tau^+ \tau^-$ \\
\hline
ggF  &  &  & & & & &   \\
$\gamma \gamma$  & 0.06 & -0.11 &  & & & &  \\   
$Z Z^*$  & & & -0.21 & & & -0.28& \\
$W^+ W^-$  &  &  & & -0.08 & &  \\
$\tau^+ \tau^-$   &  &  & &  & -0.45  & &  \\
\hline
VBF  &  &  & & & &   \\
$\gamma \gamma$  & & & 0.07   & & & \\
\hline
$VH$  &  &  & & & &   \\
$ZZ^*$   &  & & & & &   &-0.07  & \\
\hline
$t\bar{t}H$  &  &  & & & & &  \\
$W^+ W^-$  &  &  & & & & & &  -0.42 \\

\end{tabular}
\caption{Correlation matrix of the signal strength measurements used in
  the fit. Only elements larger or equal to 0.05 are used in the
  present fit and are shown in the table. The correlation matrix is
  symmetric therefore only its upper triangle section is
  reported.}
\label{tab:correlations}
\end{table}

The determination of constraints on $\ktre$ has been
performed by the ATLAS collaboration in ref.\cite{ATLAS_HH_comb}
using data from the $pp \to HH$ production measurements, with the $HH$
pair decaying to the final states $ b \bar{b}\gamma \gamma$, $b \bar{b} \tau^+
\tau^-$ and $b\bar{b} b\bar{b}$,
finding  \mbox{$ -5.0 < \ktre < 12.0 $} (\mbox{$ -5.8 < \ktre < 12.0 $})
at $95\%$ CL in observation (expectation).
In that paper only the $HH$
production cross section was parameterised as a function of
$\ktre$ while the decay branching fractions were assumed to
be independent from $\ktre$ and equal to their SM
expectation. This assumption was removed in a conference note of the
same collaboration \cite{ATLAS_H_HH_comb} where the $HH$ constraints
were combined with single-Higgs differential measurements. In the $pp
\to HH$ process the $\ktre$ value has a big impact on the
dynamic of the decay products affecting strongly the Higgs-pair
invariant mass distribution, therefore it is not possible to extract sensible
information from the final result without the information on the
likelihood of each channel expressed as a function of
$\ktre$.

\begin{figure}[tb]
\begin{center}
\includegraphics[width=0.65\textwidth]{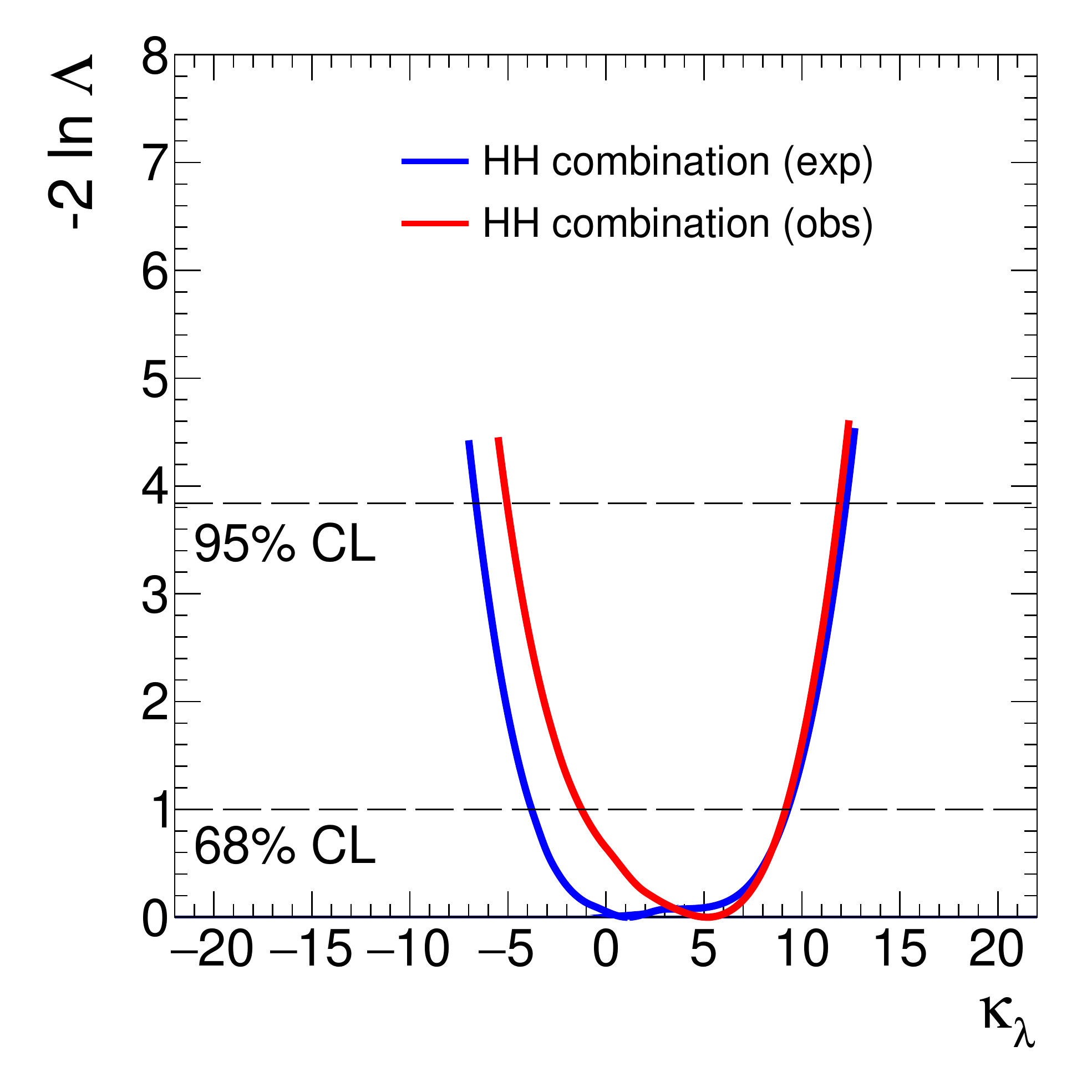}
\end{center}
\caption{Likelihood function used in the fit procedure from the
  combination of the $HH \to b\bar{b}b\bar{b}$, the $HH \to b \bar{b}  \gamma
  \gamma $ and the $HH \to b\bar{b}\tau^+\tau^-$ channels.
   } \label{fig:likelihood_interp}
\end{figure}

The likelihood shapes have been taken from
ref.\cite{EleThesis, EleProceedings}. The shape of the expected and observed
likelihood
of the combination $HH \to b\bar{b}b\bar{b}$, $HH \to 
b\bar{b} \gamma \gamma$ and $HH \to b \bar{b}\tau^+ \tau^- $ has been first scanned
with 200 points and then interpolated using a third degree
polynomial. Continuity of the first and the second derivative has been
imposed at each point of the scan.  The resulting likelihood function
used in the fit procedure is shown in
Figure~\ref{fig:likelihood_interp}.

\section{Fit procedure}
\label{sec:fit}
The fit procedure is performed by building up a likelihood function as a product of the likelihood function associated to each experimental measurement:
\[
\mathcal{L} = \mathcal{L}_{H} \times \mathcal{L}_{HH} \times \mathcal{L}_{m_W}
\times \mathcal{L}_{\sin^2 \theta^{\rm lep}_{\rm eff}}
\]
The likelihood $\mathcal{L}$ is a function of one parameter of
interest $\ktre$, the ratio: $\Lambda(\ktre) =
\mathcal{L}(\ktre)/\mathcal{L}(\hat{\kappa}_{\lambda})$
is used to extract the best fit values and confidence
intervals on $\ktre$, where $\hat{\kappa}_{\lambda}$ is the value that
maximises $\mathcal{L}$.

The likelihoods $\mathcal{L}_{H}$, $\mathcal{L}_{HH}$,
$\mathcal{L}_{m_W}$ and $\mathcal{L}_{\sineff}$ are relative to the single Higgs
production and decay
measurements, the $HH$ production, the $\mw$ and the $\sineff$ measurements
respectively.  The minimisation is performed on the quantity
$-2\,{\rm ln} \Lambda$ whose expression is:
\[
-2\,{\rm ln} \Lambda = -2 \,{\rm ln}\mathcal{L}_H -2\, {\rm ln} \mathcal{L}_{HH} -
2\, {\rm ln} \mathcal{L}_{\mw} -2 \, {\rm ln} \mathcal{L}_{\sineff} +
2\, {\rm ln}\mathcal{L}(\hat{\kappa}_{\lambda})  ~.
\]
The $-2\,{\rm ln} \mathcal{L}_{HH}$ is obtained directly from Figure
\ref{fig:likelihood_interp}, $-2\,{\rm ln}{\mathcal{L}_{m_W}}$ and
$-2\,{\rm ln}\mathcal{L}_{\sineff}$ are
$\chi^2$ function built as:
\[
-2\,{\rm ln}{\mathcal{L}_{\mw}} = \chi^2_{\mw} =
\frac{[\mw^{\rm exp} - \mw^{\rm theo}(\ktre)]^2}{[\sigma^{2 \;  exp}_{\mw} +
    \sigma^{2 \, theo}_{\mw}]^2}
\]
\[
-2\, {\rm ln}{\mathcal{L}_{\sineff}} = \chi^2_{\sineff} =
  \frac{\left[ \sin^2 \theta_{\rm eff}^{\rm lep, exp} -
      \sin^2 \theta_{\rm eff}^{\rm lep, theo}(\ktre) \right]^2}{
    \left[ \sigma^{2 \; {\rm exp}}_{\sineff} +
     \sigma^{2 \; {\rm theo}}_{\sineff} \right]^2}~.
\]
In this expression the labels "exp" and "theo" refer to the
experimental and theoretical quantities respectively, while $\sigma$
is the uncertainty on the observable denoted at the subscript. The
experimental uncertainties are listed in Table \ref{tab:input_summary}
while theoretical uncertainties are discussed in section
\ref{sec:theory}.
 
The likelihood function $\mathcal{L}_H$ contains information from
single Higgs boson production and decay measurements, the quantity $-2\,
{\rm ln} \mathcal{L}_H$ is a $\chi^2$ function defined as:
 \[
 -2\, {\rm ln} \mathcal{L}_H = \chi^2_H =
 \left[ \vec{\mu}^{\rm exp} - \vec{\mu}^{\rm theo}(\ktre) \right]^T C(\ktre)^{-1}
 \left[ \vec{\mu}^{\rm exp} - \vec{\mu}^{\rm theo}(\ktre) \right]
\]
where $\vec{\mu}^{\rm exp}$ is a fifteen dimensional vector containing
the measurements $\mu_i \times \mu_f$ listed in Table \ref{tab:mus}
and their theoretical expectation as a function of $\ktre$ described
in section \ref{sec:theory}. The matrix $C(\ktre)^{-1}$ is the inverse
of the covariance matrix $C(\ktre) = C^{\rm theo}(\ktre) +
C^{\rm exp}$ where $C^{\rm exp}$ is built from the uncertainties shown in
Table \ref{tab:mus} and the correlation matrix $\rho$ shown in Table
\ref{tab:correlations}, while $C^{\rm theo}(\ktre)$ is a diagonal
matrix containing the square of the theoretical uncertainties on
$\mu_i \times \mu_f$ due to missing higher order terms as discussed in section \ref{sec:theory}.

\section{Results}
\label{sec:result}
\begin{figure}[htbp]
    \centering
    \includegraphics[width=0.49\textwidth]{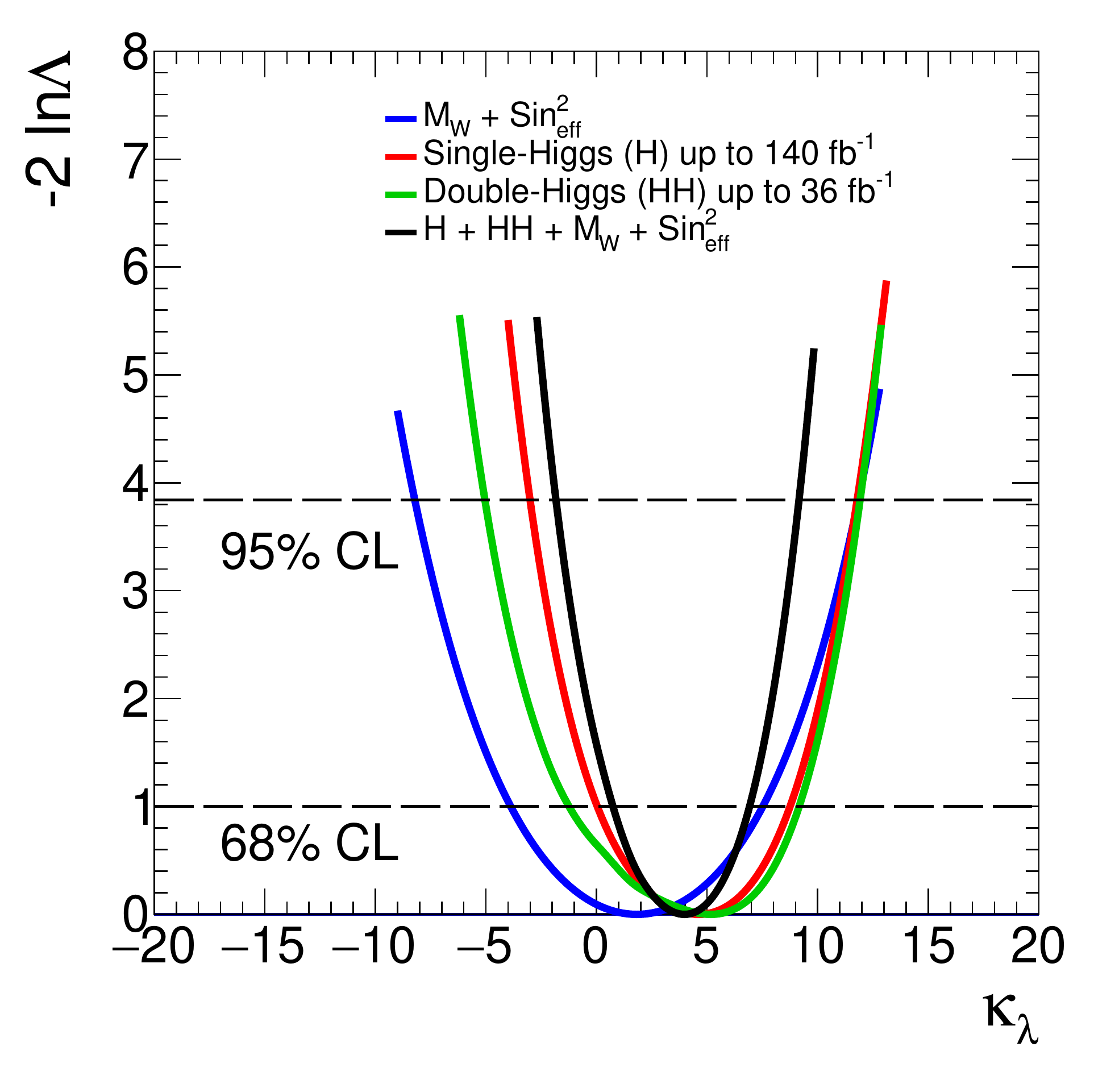}
     \includegraphics[width=0.49\textwidth]{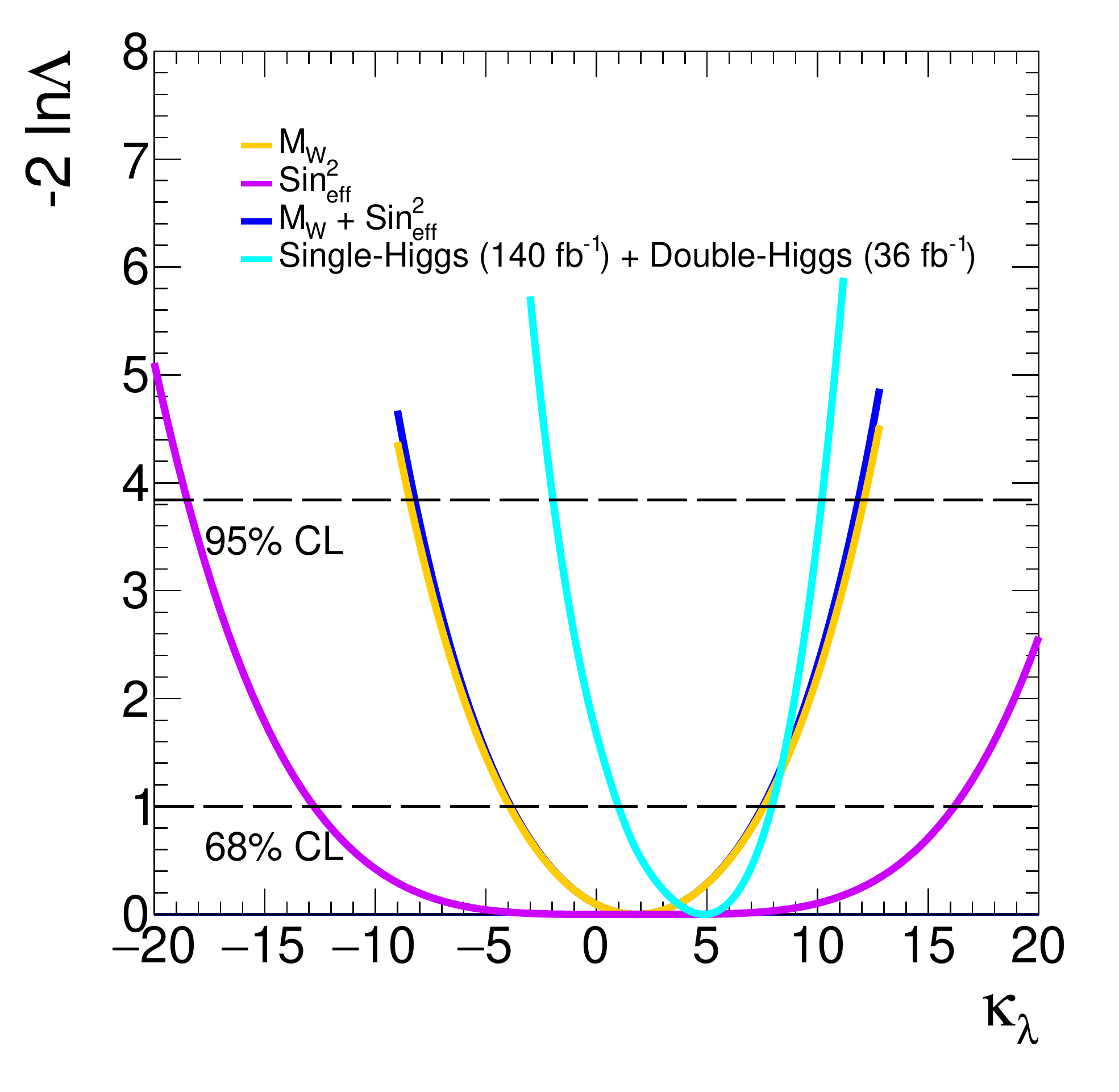}
     \caption{Values of $-2\,\rm{ln}\Lambda$ as a function of $\ktre$.
      Left panel: the $\mw + \sineff$ combination, the
      single-Higgs analyses, the double-Higgs analyses and their
      combination. Right panel: The $\mw$, the
      $\sineff$, their
      combination and the combination of the single-Higgs and
      double-Higgs analyses.}
    \label{fig:likelihood}
\end{figure}

The value of $-2\,{\rm ln}\Lambda$ as a function of the $\ktre$
parameter is shown in the left panel of Figure
\ref{fig:likelihood}. For positive $\ktre$ values, i.e. when the
interference between the box and the self-coupling diagram in the $pp
\to HH$ process is destructive and brings to a sensitivity loss of the
double-Higgs channel, all the three measurements ($HH$, single-$H$ and
EWPO) show a comparable constraining
power, with a stronger impact of the EWPO
for low values of $\ktre$. On the other hand, for negative $\ktre$
values, the higher statistics of the single-Higgs analyses allows to
reach a better constraint on $\ktre$ while the EWPO have a smaller 
impact on the result. The fit results are
summarised in Table \ref{tab:results}.

\begin{table}[t]
\centering
\begin{tabular}{lccc}
observables & best fit & 68 \% CL interval & 95 \% CL interval \\
\hline
 $\sineff$ & 0.2 & -12.8 $-$ 16.2 & -18.5 $- {} [>20]$ \\
$m_W$ & 1.8 & -3.9 $-$ 7.6 & -8.4 $-$ 12.1 \\
\hline
$\mw + \sineff$ & 1.8 &    -3.9 $-$ +7.5     & -8.2 $-$ 11.8 \\
$HH$ & 5.2 & -1.2 $-$ +9.2 & -5.0 $-$ 11.9 \\
single-$H$ & 4.6 & +0.05 $-$ +8.8 & -3.0 $-$ 11.8 \\
\hline
Combination & 4.0 & 0.7 $-$ 6.9 & -1.8 $-$ 9.2 \\
    \end{tabular}
\caption{Best fit results, 68\% and 95\% CL intervals for all measurements
  used in this work and their combination.}
    \label{tab:results}
\end{table}

\begin{figure}[t]
    \centering
    \includegraphics[width=0.49\textwidth]{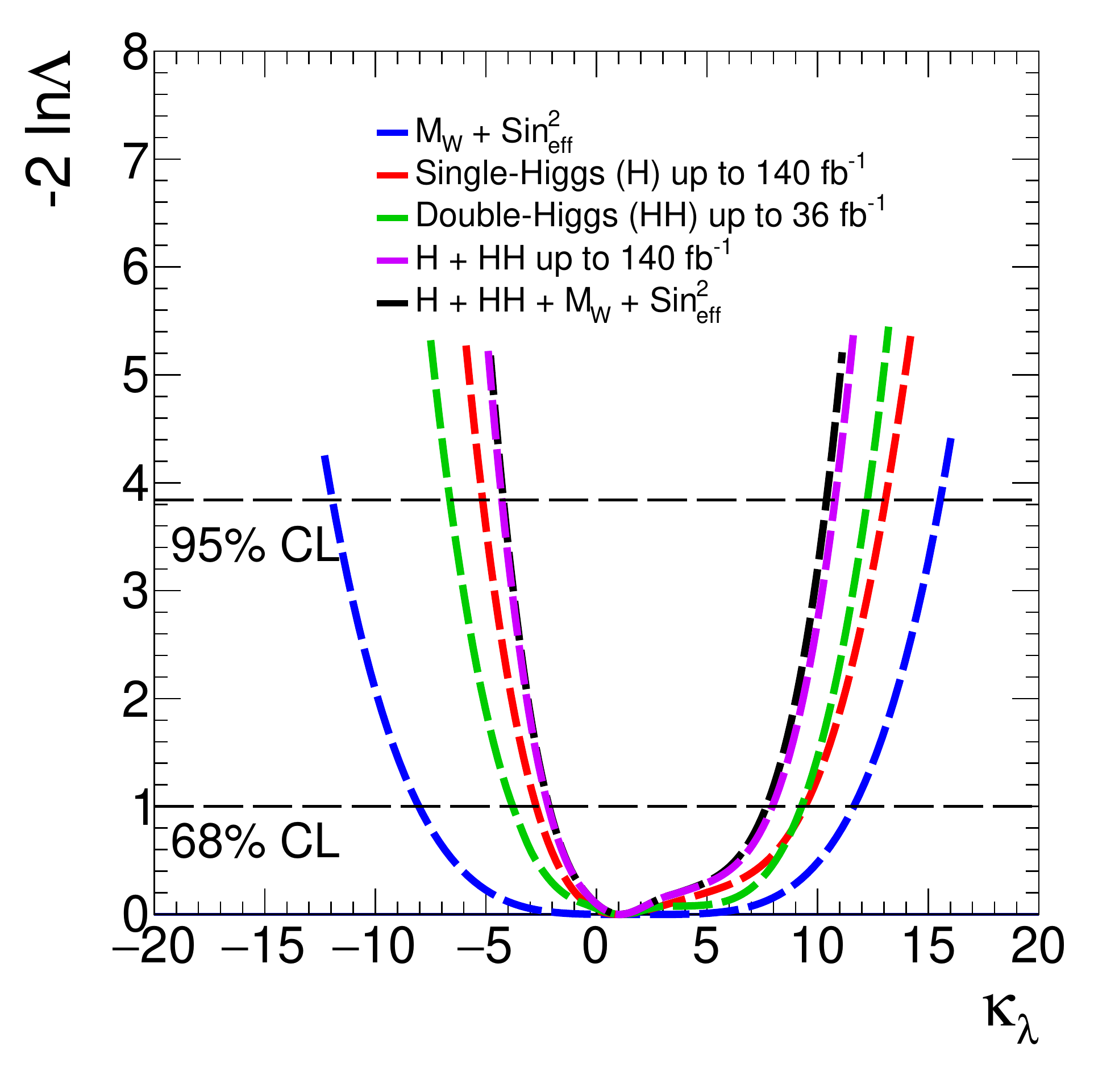}
    \caption{Value of -2\, ln$\Lambda$ as a function of $\ktre$ obtained
      setting all observables to their SM expectation. The function is
      shown for the EWPO, the
      single-Higgs, the double-Higgs observables, the single-Higgs
      plus double-Higgs combination and the full combination. }
    \label{fig:likelihood_expected}
\end{figure}

In order to compare the impact on the fit of the two  EWPO
we disentangle the likelihood functions of $\mw$ and $\sineff$ from the
$\mw + \sineff$ combination and from the combination of single-Higgs plus
double-Higgs results, as shown on the right panel of Figure
\ref{fig:likelihood}. The sensitivity of the EWPO is dominated by the 
$\mw$ measurement that represents an
important addition to the single-Higgs and double-Higgs combination.
In order to investigate if this result is due to the intrinsic
sensitivity of the EWPO, we have
performed the likelihood scan setting all the fit parameters to their
SM expectations. For the $HH$ analyses the expected likelihood
function has been taken directly from ref.\cite{EleProceedings},
while for the single-Higgs analyses and the EWPO,
it has been assumed that the correlation matrix and the fractional
error on the fitted parameters don't change when the parameters move
from their observed values to their expected ones. The resulting
-2\,ln$\Lambda$ functions are shown in Figure
\ref{fig:likelihood_expected}.

The functional shapes of -2\,ln$\Lambda$ show that the constraining
power of the EWPO is expected to be lower than what
observed in data, in fact the full combined -2\,ln$\Lambda$ doesn't show
large differences with respect to the combination of only the
single-Higgs and double Higgs -2\,ln$\Lambda$.  From Figure
\ref{fig:likelihood} is possible to see that the combined $H$ and $HH$
-2\,ln$\Lambda$ has a minimum far from its SM expectation of
$\kappa_\lambda = 1$, while the minimum of the EWPO
-2\,ln$\Lambda$ is closer to its SM expectation. Therefore the
EWPO have an higher impact on the final observed
result, in particular at the upper edge of the confidence interval.

\section{Conclusion}
\label{sec:conc}
In this paper we combine the ATLAS data analyses of the 
single-Higgs and double-Higgs processes with the information coming
from the EWPO in order to constraint the Higgs boson trilinear self-coupling
modifier $\ktre = \tril/\trilsm$. Under the assumption that NP affects only
the Higgs potential we find as the best fit value of the trilinear
self-coupling modifier $\ktre= 4.0^{+2.9}_{-3.3}$ excluding values outside
the interval \mbox{$ -1.8 < \ktre < 9.2 $} at $95\%$ CL. 
With respect to analyses
where single-Higgs data \cite{ATLAS_H_HH_comb} or double-Higgs
data \cite{ATLAS_HH_comb} or a combination of both \cite{ATLAS_H_HH_comb}
are taken into account, our study shows that the inclusion in the fit of
the information coming from the EWPO $\mw$ and $\sineff$ gives rise to a
stronger constraint on $\ktre$, in particular on the positive side of the
CL interval.

At the moment the information coming from EWPO gives an indication for $\tril$
values closer to $\trilsm$ than the single and double-Higgs analyses. It is
interesting to see if, in the future, with the LHC collaborations analysing larger set
of single and double-Higgs data and with possible improvements on the measurement
of the $\mw$ from LHC, this different indication will remain in the data.

\section*{Acknowledgements}
The work of G.D. was partially supported by the Italian Ministry of Research (MUR) under grant PRIN 20172LNEEZ.
The work of PPG has received financial support from Xunta de Galicia (Centro singular de investigaci\'on de Galicia accreditation 2019-2022), by European Union ERDF, and by  ``Mar\'ia  de Maeztu"  Units  of  Excellence program  MDM-2016-0692  and  the Spanish Research State Agency.

\end{document}